\documentclass{rmf-d}
\usepackage{nopageno,rmfbib,multicol,times,epsf,amsmath,amssymb,cite}
\usepackage[latin1]{inputenc}
\usepackage[]{caption2}
\usepackage{graphics}
\usepackage{graphicx}
\usepackage[hidelinks]{hyperref}
\hypersetup{colorlinks, citecolor=black, linkcolor=black ,urlcolor={blue}}
\usepackage{wrapfig, blindtext}
\usepackage{comment}
\usepackage{tabularx}
\usepackage{float}
\usepackage{dblfnote}

\newcommand{\R}{\mathbb{R}}
\newcommand{\C}{{\kern+.25em\sf{C}\kern-.45em\sf{{\small{I}}} \kern+.45em\kern-.25em}}

\newcommand{\be}{\begin{equation}}
\newcommand{\ee}{\end{equation}}
\newcommand{\bea}{\begin{eqnarray}}
\newcommand{\eea}{\end{eqnarray}}
\newcommand{\nn}{\nonumber}

\newcommand{\la}{\langle}
\newcommand{\ra}{\rangle}
\def\rmfcornisa{ \hfill\rmf\ {\bf }
\hfill }

\def\rmfcintilla{{\it Rev.\ Mex.\ Fis.\/} {\bf } }
\clearpage \rmfcaptionstyle 
\begin{document}
\title{Semi-vortices and cluster-vorticity: new concepts in the Berezinski\u{\i}-Kosterlitz-Thouless phase transition
\vspace{-6pt}}
\author{B.\ G\'{o}mez Bravo, B.~D.\ Ju\'{a}rez Hern\'{a}ndez, W.\ Bietenholz}
\address{Instituto de Ciencias Nucleares, Universidad Nacional Aut\'{o}noma
de M\'{e}xico \\ A.P.\ 70-543, C.P.\ 04510 Ciudad de M\'{e}xico, M\'{e}xico}
\maketitle
\begin{abstract}
\vspace{1em}
The Berezinski\u{\i}-Kosterlitz-Thouless (BKT) essential phase
transition in the 2d XY model is revisited. Its mechanism
is usually described by the (un)binding of vortex--anti-vortex
(V--AV) pairs, which does, however, not provide a clear-cut quantitative
criterion for criticality. Known sharp criteria are the
divergence of the correlation length and a discontinuity of the
helicity modulus. Here we propose and probe
a new criterion: it is based on the concepts of semi-vortices
and cluster vorticity, which are formulated in the framework of the
multi-cluster algorithm that we use to simulate the 2d XY model.
\vspace{1em}
\end{abstract}
\keys{2d XY model, essential phase transition, vortices, cluster algorithm \vspace{-4pt}}
\pacs{05.10.Ln, 5.10.Hk, 11.10.Kk, 64.60.De, 64.70.-p \vspace{-4pt}}
\begin{multicols}{2}

\section{The 2d XY model}
  
We consider the 2d XY model, or 2d O(2) model, on square lattices
of size $L \times L$. At each lattice site $x=(i,j)$,
$i,j \in \{1, \dots ,L \}$ there is a classical spin variable
$\vec e_{x} \in S^{1}$, {\it i.e.}\ $\vec e_{x} \in \R^2$ and
$|\vec e_{x}|=1, \ \forall x$. It can be parameterized as
$\vec e_{x} = (\cos \varphi_{x}, \sin \varphi_{x})$, $\varphi_{x} \in \R$.
In lattice units, the Hamilton function, or Hamiltonian, of a spin
configuration $[\vec e \, ]$ is given by
\be  \label{Hami}
{\cal H}[\vec e \, ] = - \sum_{\la xy \ra} \vec e_{x} \cdot \vec e_{y} \ ,
\ee
where $\la xy \ra$ indicates the nearest-neighbor lattice sites; we
see that the spins are coupled
ferromagnetically.\footnote{We do not include a coupling constant,
  because what matters below is just the ratio ${\cal H}/T$, where $T$
  is the temperature, which we scale such that it absorbs this
  coupling. In lattice units (with lattice spacing 1) the dimensions
  of ${\cal H}$ and $T$ are not manifest.}
The model has a global O(2) symmetry, which inspires its application
in the description of films of superfluid $^{4}$He and of superconductors
(although there the ${\rm U}(1)= {\rm O}(2)$ symmetry is local).

We assume periodic boundary conditions in both directions,
{\it i.e.}\ the volume has the structure of a torus.
As usual, the partition function $Z$ and the thermal expectation value
of some observable $A[\vec e \, ]$ are given by the functional integrals
\be
Z = \int {\cal D}\vec e \ e^{-{\cal H}[\vec e \, ] /T} \ , \
\la A \ra = \frac{1}{Z} \int {\cal D}\vec e \ A[\vec e \, ] \nn
e^{-{\cal H}[\vec e \, ] /T} \ ,
\ee
where $T$ is the temperature, cf.\ footnote 1.

Of particular interest are {\em vortices} in the spin configurations.
In order to define them, we consider the relative angle
between two nearest-neighbor spins,
\be
\Delta \varphi_{x, x + \hat \mu} =  (\varphi_{x + \hat \mu} - \varphi_{x})
\ {\rm mod} \ 2\pi \in (-\pi , \pi ) \ ,
\ee
where $\hat \mu$ is a lattice unit vector in the $\mu$-direction.
Note that we are using a non-standard modulo operation,
which minimizes the absolute value (the ambiguous case
$\Delta \varphi_{x, x + \hat \mu } = \pm \pi$ has measure zero).
Each plaquette carries a vorticity number
\bea
v_{x} \!\!\!&=&\!\!\! \frac{1}{2\pi} \left( \Delta \varphi_{x, x + \hat 1} +
\Delta \varphi_{x + \hat 1, x + \hat 1 + \hat 2} +
\Delta \varphi_{x + \hat 1 + \hat 2, x + \hat 2} \right. \nn \\
&&\!\!\! \left. + \Delta \varphi_{x + \hat 2, x} \right) \in \{ 1, 0,-1 \} \ .
\eea
$v_{x} = 1$ means that a vortex (V) is located on this plaquette,
for $v_{x} = -1$ it is an anti-vortex (AV), and for $v_{x} = 0$
the plaquette is neutral (free of vorticity).
\begin{figure}[H]
\centering
\includegraphics[scale=0.3]{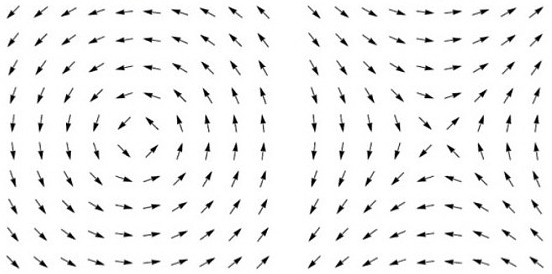}
\caption{Prototypes of a vortex (V, left) and an anti-vortex (AV, right) in
  the centers of the plots. They show subsets of possible configurations
  in some region.}
\label{VandAV}
\vspace*{-2mm}
\end{figure}

Due to Stokes' Theorem, the periodic boundary conditions imply that the
total vorticity of any configuration vanishes, $\sum_{x} v_{x} = 0$,
hence there is always the same number of V and AV.

At the critical temperature $T_{\rm c}$ this model undergoes an essential
phase transition, the {\em Berezinski\u{\i}-Kosterlitz-Thouless
(BKT) transition} \cite{VLB,KT}. For a comprehensive overview, we refer to
Ref.\ \cite{40years}. BKT transitions have been experimentally observed
in superfluid $^{4}$He \cite{superfluid}, in various superconductors
\cite{superconductor}, and recently also in a frustrated magnet \cite{Nature}.

In contrast to finite-order
phase transitions, the correlation length $\xi$ does not diverge
with a power-law, $\xi \propto (T-T_{\rm c})^{- \nu}$, but its behavior at
$T \gtrsim T_{\rm c}$ is exponential,
\be
\xi \sim \exp \left( \frac{\rm const.}{(T - T_{\rm c})^{\nu_{\rm e}}} \right) \ ,
\ee
where $\nu_{\rm e} = 0.5$ \cite{Kos74} is an exponential critical
exponent.\footnote{Here the relation $\sim$ means that this is the
  term which dominates the divergence, even if there may be a
  pre-factor with some power of $(T - T_{\rm c})$.}
In Ehrenfest's scheme, this is a phase transition of infinite order,
{\it i.e.}\ very smooth. It is not related to any spontaneous
symmetry breaking (the spontaneous breaking of the O(2) global
symmetry is excluded in $d=2$ by the
Mermin-Wagner-Coleman Theorem,
so we avoid the expressions ``order'' and ``disorder'').
At $T<T_{\rm c}$ the system remains critical, covering a variety of
universality classes.

The critical temperature was numerically measured in numerous works.
High-precision results were reported in Refs.\ \cite{Has05}, which are
in agreement with $\beta_{\rm c} \equiv 1/T_{\rm c} = 1.1199(1)$.

In a sequence of famous papers \cite{VLB,KT}, Berezinski\u{\i}
and later Kosterlitz and Thouless assigned this transition to the
vortex dynamics:
\begin{itemize}

\item At $T< T_{\rm c}$ the V and AV appear
in pairs close to each other (``bound pairs''): for a given number
of V and AV, this structure minimizes the free energy $F = -T \ln Z$,
which implies an attractive force between near-by V and AV. Such localized
pairs do not significantly affect the long-range correlations, so we are
in the {\em massless phase}.

\item At $T>T_{\rm c}$ these pairs ``unbind'' as an entropy effect:
  the attractive force loses its dominance over the high multiplicity
  of configurations where V and AV are spread over the volume without
  any specific structure.\footnote{Schematically we see this by writing
    the free energy as $F = E_{\rm V} - TS = (\pi -2T) \ln L$, where
    $E_{\rm V}$ is the (estimated) energy requirement for inserting one
    V or AV into a ``smooth background'', and $S$ is the entropy, as
    reviewed in Ref.\ \cite{WBUrs}.}
  We denote them as ``free'' V and AV, and their significant density
  does affect long-range correlations, which entails the
  {\em massive phase}.
   
\end{itemize}
\begin{figure}[H]
\centering
\vspace*{-16mm}
\hspace*{-4mm}
  \includegraphics[scale=0.216]{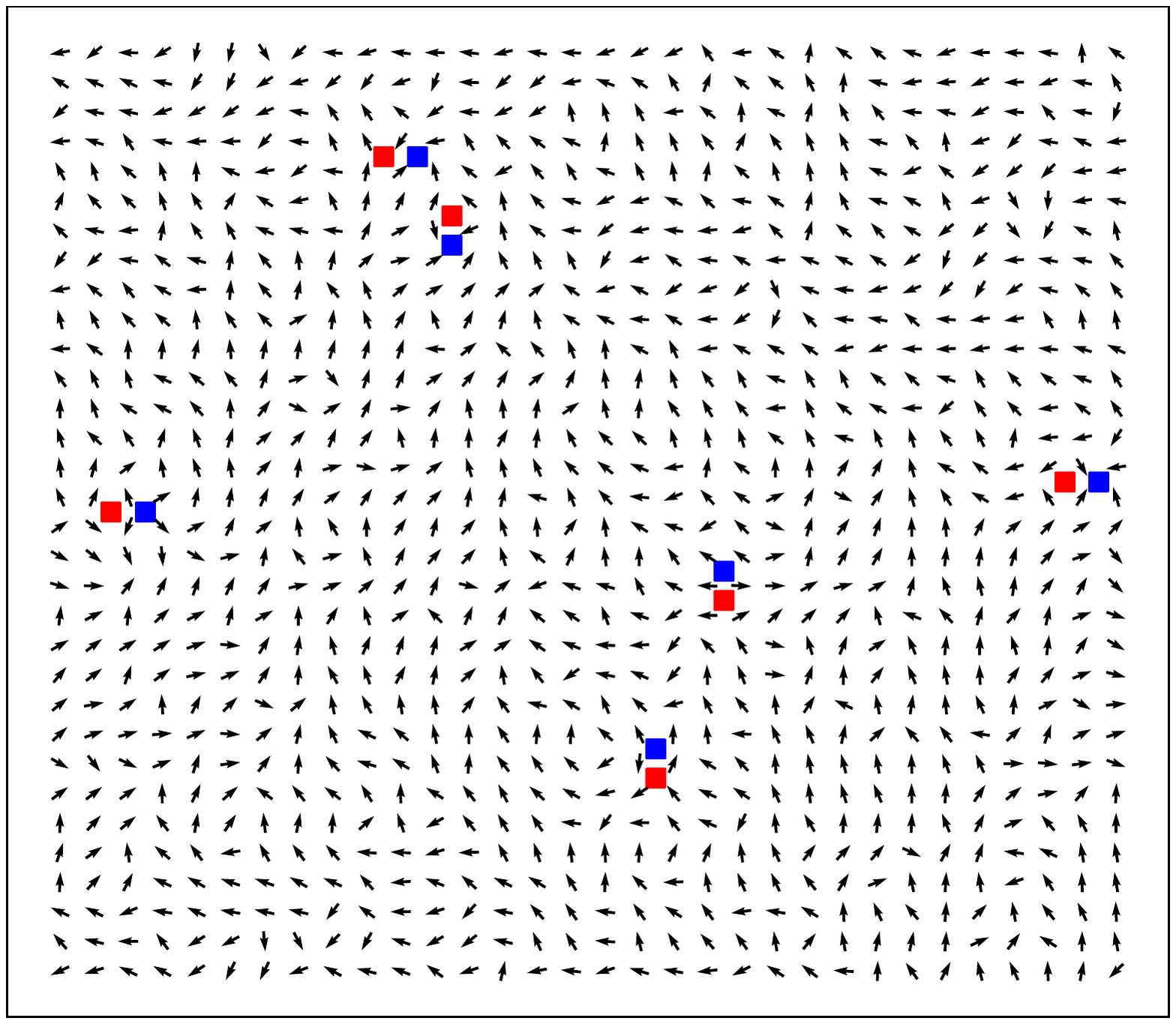}
  \hspace*{-8mm}
  \includegraphics[scale=0.216]{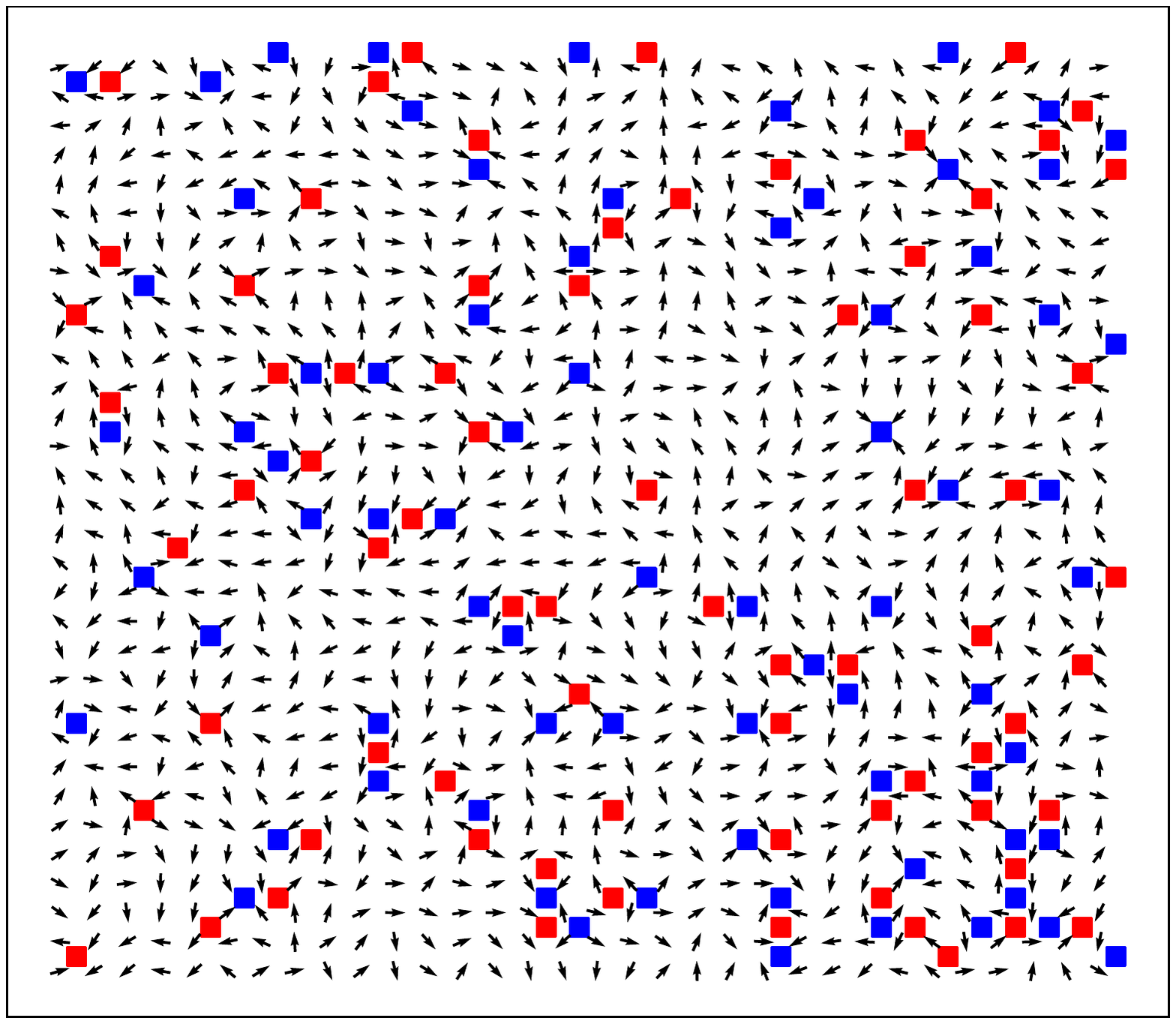}
 \vspace*{-13mm}
 \caption{Typical configurations at $\beta = 1.12 > \beta_{\rm c}$ (left)
   and at $\beta = 0.6 < \beta_{\rm c}$ (right), with V and AV indicated by
   red (bright) and blue (dark) plaquettes, respectively (and
   $\beta \equiv 1/T$). We see bound
   V--AV pairs on the left, versus free V and AV on the right.}
\label{VAVpairs}
\vspace*{-2.7mm}
\end{figure}

In 2016 Kosterlitz and Thouless were awarded the Nobel Prize
(sadly Berezinski\u{\i}\footnote{His work particularly
  inspired Polyakov to introduce his famous ``dislocations'' in gauge
  theory, with the hope to explain the
  confinement--deconfinement transition \cite{Poly}.}
  had passed away in 1980, at the
age of only 44) --- at that occasion, their work was reviewed
in Ref.\ \cite{WBUrs}.

\vspace*{-1mm}
\subsection{Constraint lattice Hamiltonian}

Thanks to {\em universality,} this model --- like other models ---
can be formulated with an (infinite) variety of lattice Hamiltonians.
The continuum limit --- which corresponds to $\xi \to \infty$,
so that the lattice spacing vanishes in units of $\xi$ as an
intrinsic scale --- is always the same (if suitable conditions
are fulfilled, such as locality), hence they all describe the
same physics.

Constraint lattice Hamiltonians for O($N$) models have a simple
structure, without any derivatives \cite{topact}: there is just
a constraint $\delta$ for all angles between nearest-neighbor spins,
\be
   {\cal H}[\vec e \, ] = \left\{ \begin{array}{ccc}
     0 && {\rm if} ~ | \Delta \varphi_{x,x+\hat \mu} | < \delta, \quad
     \forall x, \mu \\
     + \infty && {\rm otherwise.} \\
     \end{array} \right.
\ee
Instead of temperature one varies the constraint angle $\delta$,
and the BKT transition occurs in the 2d XY model
at $\delta_{\rm c} = 1.775(1)$ \cite{topactXY}. We confirmed
that it is in the BKT universality class by numerically measuring the
exponential critical exponent $\nu_{\rm e} =  0.503(7)$, which is compatible
with Kosterlitz' prediction \cite{Kos74}. Further evidence for the BKT
universality class is based on the step-scaling function
\cite{Balog,topactXY}.

In Ref.\ \cite{constraint} we studied in particular the (un)binding
mechanism, by considering various options for a cutoff distance between
the nearest V and AV, for being considered a ``bound pairs''.
Indeed, the celebrated mechanism is confirmed again.
However, the argument with the free energy does not apply in this
formulation, which does not really agree with the established picture:
the (un)binding mechanism seems to be a pure entropy effect after all.

\vspace*{-1mm}
\subsection{Helicity modulus}

A short-coming of the (un)binding mechanism is that
it does not provide a clear-cut quantitative criterion
for $T_{\rm c}$. In particular, the distinction between
``bound pairs'' and ``free'' V and AV is not clearly defined.

\begin{figure}[H]
\centering
  \vspace*{-2mm} 
  \includegraphics[scale=0.55]{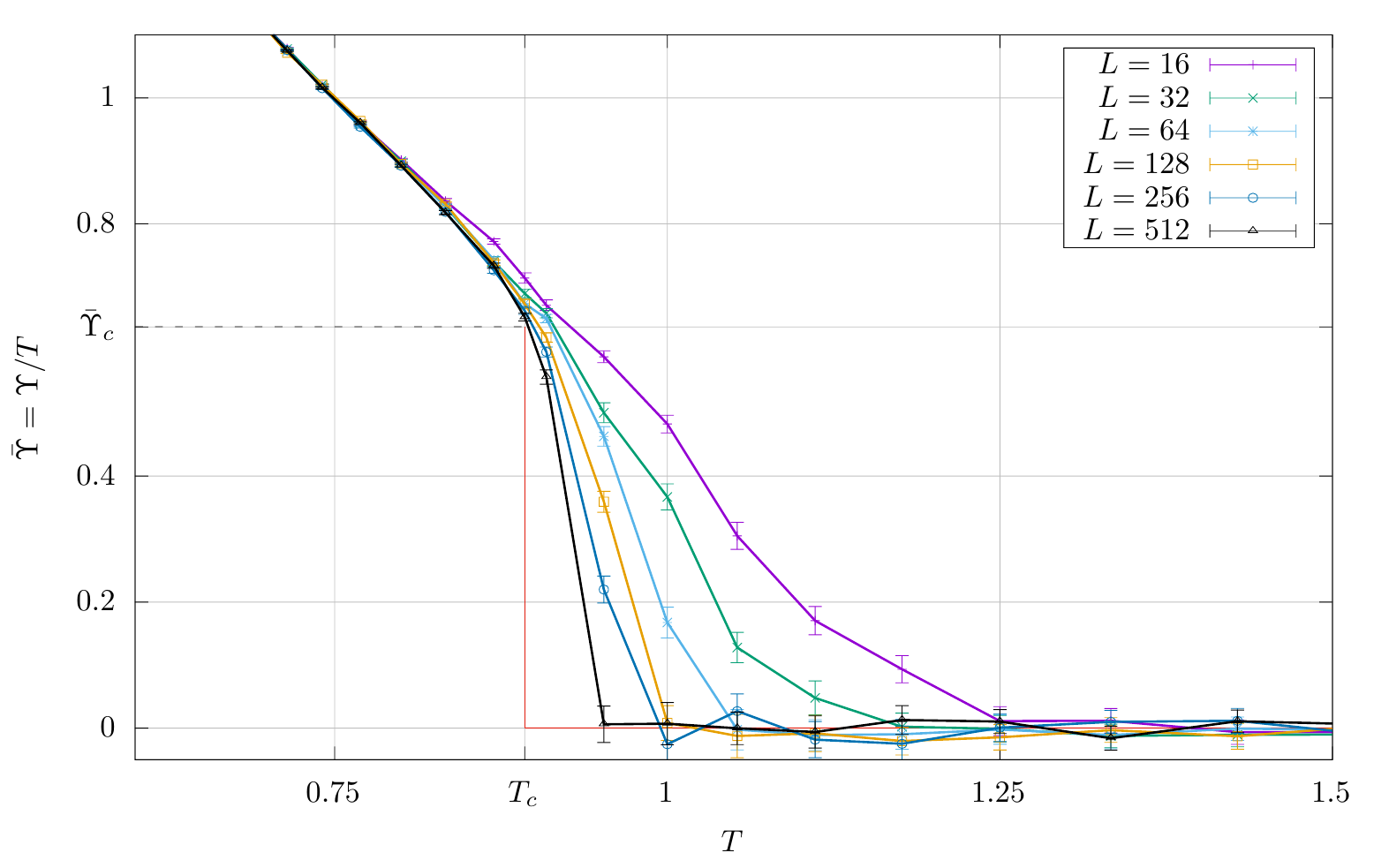}
 \vspace*{-4mm}
 \caption{Simulation results for the
  dimensionless helicity modulus $\bar \Upsilon$, obtained with the
  standard Hamiltonian (\ref{Hami}), and the theoretically predicted
  jump in infinite volume.}
\label{helimod}
\vspace*{-4mm}
\end{figure}

We have mentioned the divergence of the correlation length
for $T \searrow T_{\rm c}$ as one clear criterion. Another one refers
to the {\em helicity modulus} (or {\em spin stiffness}):
it quantifies how the free energy $F$ reacts to an (infinitesimal)
change of the boundary conditions. Say in the $\hat 1$-direction they
are generalized from periodic to twisted, 
\be
\vec e_{x + L \hat 1} = \left( \begin{array}{cc} \cos \alpha & -\sin \alpha \\
\sin \alpha & ~~~\cos \alpha \end{array} \right) \vec e_{x} \ .
\ee
In terms of the {\em twist angle} $\alpha$, the helicity modulus $\Upsilon$
--- and its dimensionless counterpart $\bar \Upsilon$ --- are given by
\be
\Upsilon := \frac{\partial^{2} F}{\partial \alpha^{2}} \ , \quad
\bar \Upsilon := \frac{1}{T} \Upsilon \ .
\ee
For the standard Hamiltonian (\ref{Hami}) one obtains\footnote{There is
  some confusion about the correct term, but it has been reproduced
  carefully in Refs.\ \cite{theses}.}
\bea
\bar \Upsilon &=& \frac{1}{TL^{2}} \Big\la
\sum_{x} \vec e_{x} \cdot \vec e_{x+\hat 1} \Big\ra \nn \\
&& - \frac{1}{(TL)^{2}} \Big\la \Big( \sum_{x} (e_{x}^{(1)} e_{x+ \hat 1}^{(2)}
- e_{x}^{(2)} e_{x+ \hat 1}^{(1)} ) \Big)^{2} \Big\ra \ , \quad
\eea
hence this quantity can be numerically measured without ever moving
away from $\alpha =0$ (this is not the case for the constraint
Hamiltonian \cite{constraint}).
Theory predicts a discontinuous jump of $\bar \Upsilon$ \cite{NK77},
which agrees with the observed jump in the density of superfluid
$^{4}$He films \cite{superfluid} and in a trapped, ultracold 2d Bose
gas \cite{NLM},
\bea
^{\lim}_{T \nearrow T_{\rm c}} \bar \Upsilon (T) &=& \frac{2}{\pi}
(1 - 16 e^{-4\pi}) \simeq 0.6365 \ , \nn \\
\bar \Upsilon (T \geq T_{\rm c}) &=& 0
\label{jump}
\eea
(the small exponential correction to the jump height was discovered
in Ref.\ \cite{ProSvi}). This formula refers to infinite volume, and
for the standard Hamiltonian (\ref{Hami}) the convergence towards this value
for increasing $L$ is very slow: even at $L=2048$, $T=T_{\rm c}$, it is
still 6.6\% too high \cite{Has05}. The constraint Hamiltonian behaves
much better: at $L=64$, $\delta = \delta_{\rm c}$, it already agrees with
the theory to a precision below 1\% \cite{constraint}, which was the
first convincing numerical confirmation of the prediction (\ref{jump}).
In Fig.\ \ref{helimod} we show new simulation results with the standard
Hamiltonian, which are consistent with Ref.\ \cite{Has05}, though we are
limited to $L \leq 512$.

\section{Proposal for a new criticality criterion}

Let us now proceed to our suggestion for a new criterion to identify
the critical temperature $T_{\rm c}$. It is based on the Wolff cluster
algorithm \cite{Wolff}, which is not only most efficient
to simulate this model, but it also allows us to introduce physical
quantities, like semi-vortices \cite{Lat19}. The idea is related to
the stochastic formulation of merons (semi-instantons) in the 2d O(3)
model \cite{meron}.

Let us begin by sketching the multi-cluster algorithm. A 
configuration $[\vec e \, ]$ is efficiently updated by the following
steps:

{\small
\begin{itemize}

\item 1. Choose a random vector $\vec r \in S^{1}$ with uniform probability.
  The line orthogonal to $\vec r$ through the origin is denoted as
  the Wolff line. ``Flipping'' a spin variable $\vec e_{x}$ means that
  it is reflected in the Wolff line.
  
\item 2. Check all links between nearest-neighbor lattice sites,
  say $x$ and $x + \hat \mu$ (possibly across
  the periodic boundaries). We consider putting a ``bond''
  which ``attaches'' the spin variables $\vec e_{x}$ and
  $\vec e_{x + \hat \mu}$. If we flip one of these two spins, the
  contribution of this spin pair to the Hamiltonian ${\cal H}$
  changes by some amount that we call $\Delta {\cal H}$.

  If $\Delta {\cal H} \leq 0$, we do not put a bond. In particular,
  this means that we never put a bond if $\vec e_{x}$ and
  $\vec e_{x + \hat \mu}$ point to opposite sides of the Wolff line.

  If $\Delta {\cal H} > 0$, we put a bond with probability
  $1 - e^{-\Delta {\cal H} /T}$.

\item 3. All spins which are (directly or indirectly) connected by
  bonds form one {\em cluster.} We identify all the clusters;
  thus the entire configuration is divided into a set of clusters.

\item 4. Each cluster is flipped with probability $1/2$: in this case,
  all the spins of a cluster are flipped collectively.

\item Return to 1.
  
\end{itemize}
}
\vspace*{1mm}

This algorithm fulfills the required conditions of detailed balance
and ergodicity \cite{Wolff}.
It turns out that the cluster size distribution scales with
the correlation length to some fractal dimension $D$, {\it i.e.}\
the histogram stabilizes as a function of (cluster size)$/\xi^{D}$,
thus exhibiting a universal behavior \cite{Lat19}.

Let us assume the steps 1 to 3, {\it i.e.}\ we have a ``map'' of clusters.
The V and AV can only be located at the boundaries
between the clusters; for a plaquette inside a cluster all spins point
to the same side of the Wolff line, hence these plaquettes are neutral.
It turns out that a V or AV on a plaquette can only involve spins
belonging to exactly two clusters \cite{theses,Lat19}.

This inspires the assignment of a vorticity contribution of some plaquette
to a given cluster ${\cal C}$, we call it $v_{x, {\cal C}}$.
Flipping the cluster ${\cal C}$ (while keeping all other clusters fixed)
changes the configuration $[\vec e \,]$
to another configuration $[\vec e\, ' \, ]$, and it may
change the vorticity of the plaquettes at its boundary. We define
\be
v_{x, {\cal C}} = \tfrac{1}{2} \left( v_{x}[\vec e \,] -v_{x}[\vec e\,' \,]
\right) \in \{ -\tfrac{1}{2}, 0 , \tfrac{1}{2} \} \ ,
\ee
which introduces the concept of {\em semi-vortices}
(semi-V, $v_{x, {\cal C}} = \tfrac{1}{2}$) and {\em semi-anti-vortices}
(semi-AV, $v_{x, {\cal C}} = -\tfrac{1}{2}$) \cite{Lat19}.
Note that $v_{x, {\cal C}}$ does not depend on the flipping orientation
of the other clusters, which is important for the concept to be sensible.
Thus a V is split into two semi-V associated with two clusters
(and the same for an AV). The vorticity of a plaquette
is retrieved as $v_{x} = \sum_{\cal C} v_{x, {\cal C}}$.

For a configuration with $N_{\cal C}$ clusters, all the cluster flips
provide an ensemble of 
$2^{N_{\cal C}}$ configurations. In two of them all the spins are on the
same side of the Wolff line, these are the {\em reference configurations}.
If we start from a reference configuration and flip just one cluster,
some semi-V may appear at its boundary, and the same number of semi-AV
--- they are alternatingly ordered along the boundary \cite{Lat19}.

This provides a sharp criterion to define whether or not a V--AV
pair is {\em bound:} this is the case if its semi-V and semi-AV
are associated with only two clusters. Under cluster flips, they
can only appear or disappear simultaneously, and their signs
can only change simultaneously. Here, the binding does not refer to
their distance, but to their fate under cluster flips.

\begin{figure}[H]
\centering
  \vspace*{-2mm} 
  \includegraphics[scale=0.3]{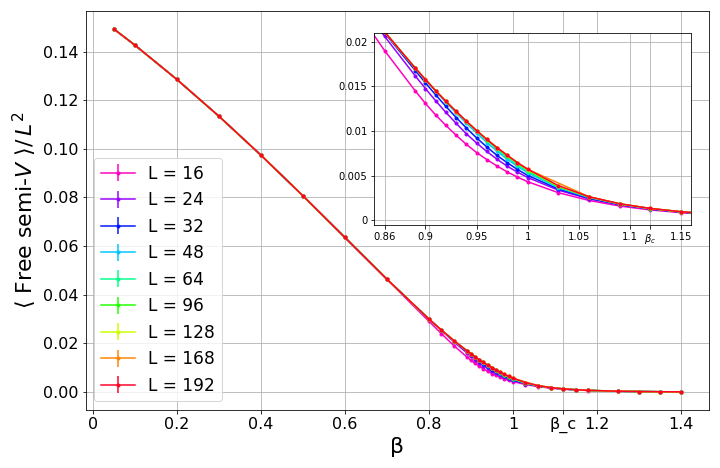}
 \vspace*{-2mm}
 \caption{The temperature-dependent density of free semi-V, as defined
   in this section, in volumes $L \times L$, with $\beta = 1/T$.}
\label{semiV}
\vspace*{-4mm}
\end{figure}

The remaining\, {\em free semi-V} are therefore supposed to drive the
BKT phase transition. Fig.\ \ref{semiV} shows that --- for decreasing
temperature $T$ --- a significant density of free semi-V does set in
around $T_{\rm c}$, but this density looks like a smooth function of $T$:
for increasing size $L$ it does not approach the behavior like an order
parameter (zero at $T \geq T_{\rm c}$, non-zero at $T < T_{\rm c}$).

\begin{figure}[H]
\centering
  \vspace*{-3mm} 
\includegraphics[scale=0.5]{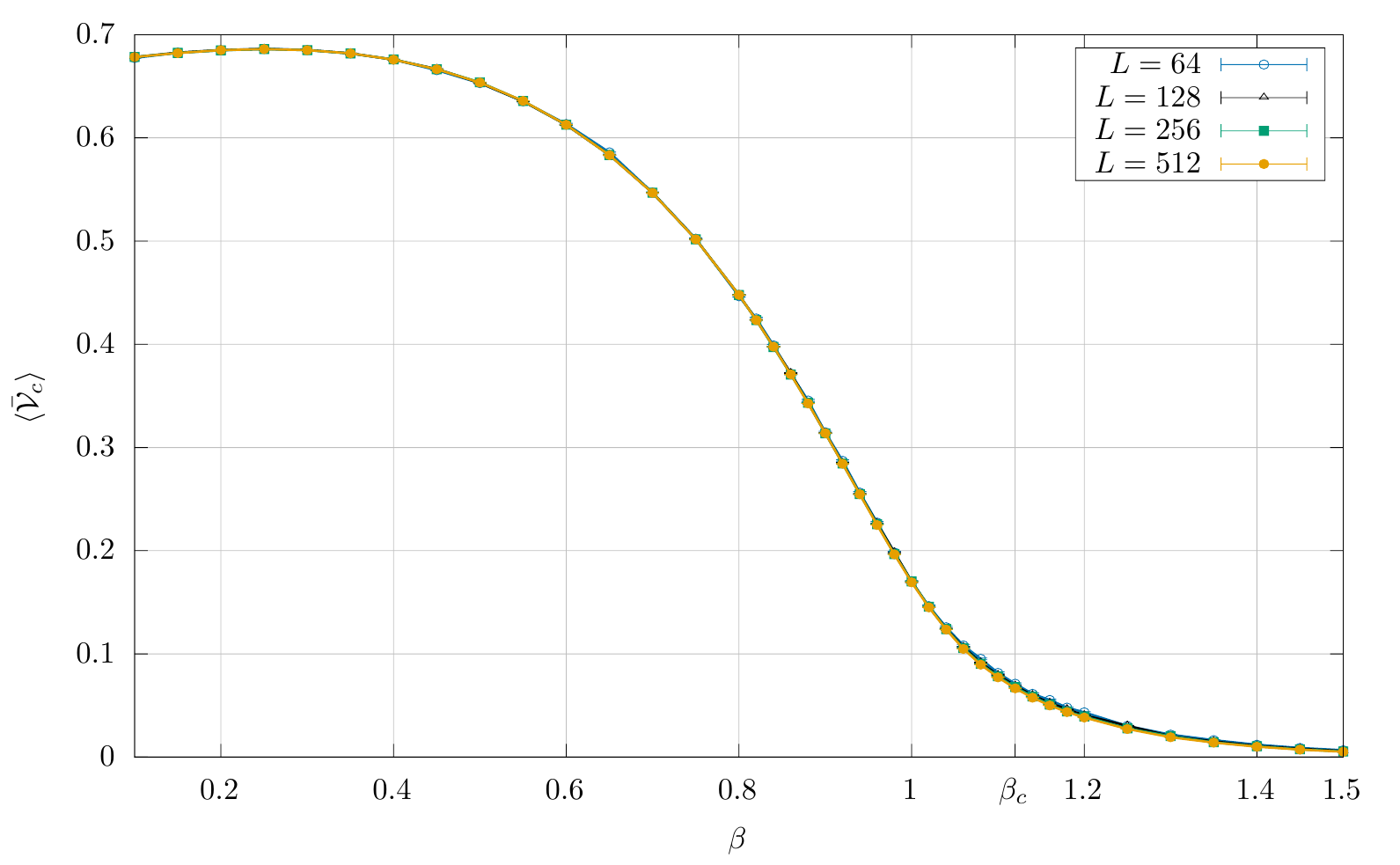}
 \vspace*{-2mm}
 \caption{Cluster vorticity $\la \bar {\cal V}_{\cal C} \ra$ as a function
   of $\beta$. It coincides to high accuracy for different lattice sizes $L$.}
\label{Cvorticity}
\vspace*{-3mm}
\end{figure}

\begin{figure}[H]
\centering
\includegraphics[scale=0.33]{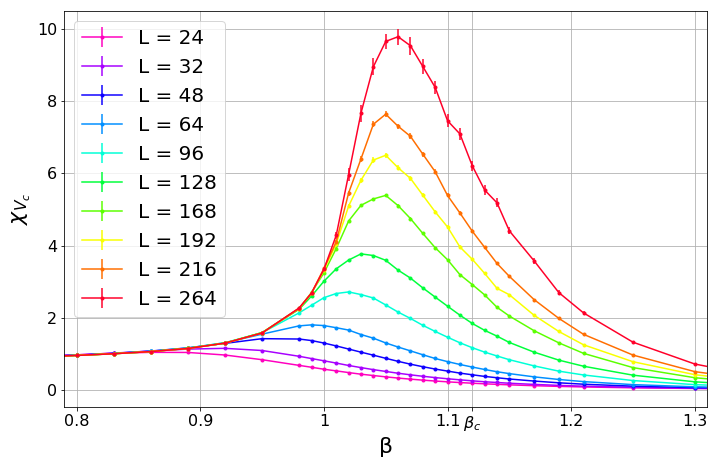}
\vspace*{-3mm}
\caption{Cluster vorticity susceptibility $\chi_{\cal V}$ as a
  function of $\beta$, for various lattice sizes $L$.}
 \label{Csus}
\end{figure}

Next we define the vorticity of a cluster ${\cal C}$ by the number of
semi-V generated by its flip out of a reference configuration,
\be
{\cal V}_{\cal C} = N_{\mbox{semi-V}} =  N_{\mbox{semi-AV}} \ .
\ee
For a configuration with $N_{\cal C}$ clusters we obtain the mean cluster
vorticity $\bar {\cal V}_{\cal C}$, and the mean square
$\bar {\cal V}_{{\cal C},2}$,
\be
\bar {\cal V}_{\cal C} = \frac{1}{N_{\cal C}} \sum_{k=1}^{N_{\cal C}}
     {\cal V}_{{\cal C}_{k}} \ , \quad
\bar {\cal V}_{{\cal C},2} = \frac{1}{N_{\cal C}} \sum_{k=1}^{N_{\cal C}}
     {\cal V}_{{\cal C}_{k}}^{2} \ .     
\ee
This takes us to the {\em cluster vorticity susceptibility}
\be  \label{clusus}
\chi_{\cal V} = \la \bar {\cal V}_{{\cal C},2} \ra
  - \la \bar {\cal V}_{\cal C} \ra^{2} \ .
\ee
Fig.\ \ref{Cvorticity} shows $\la \bar {\cal V}_{\cal C} \ra$, which
hardly depends on the volume, and which increases significantly below
$T_{\rm c}$, like the free semi-V density, but again with a smooth behavior.
The cluster vorticity susceptibility $\chi_{\cal V}$, however,
has a strong peak, if $L$ is not too small, and its height increases
with $L$, see in Fig.\ \ref{Csus}.

\begin{figure}[H]
\centering
\vspace*{-1cm}
\includegraphics[scale=0.37]{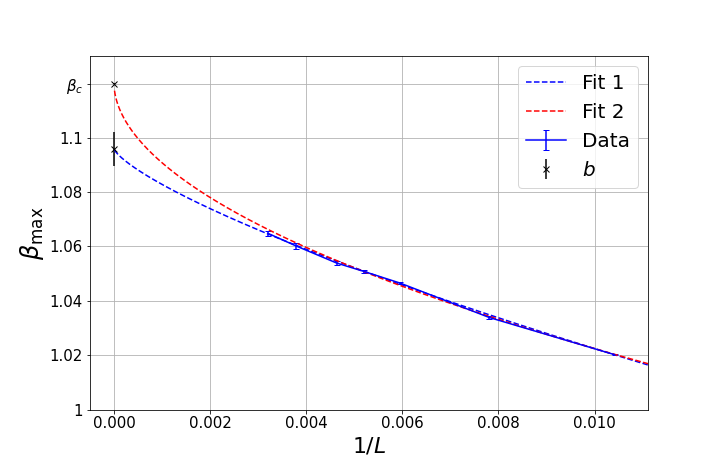}
\vspace*{-5mm}
\caption{Two thermodynamic extrapolations of $\beta_{\rm max}$,
   the peak location of $\chi_{\cal V}$.}
\label{Csusextra}
\vspace*{-3mm}
\end{figure}

Hence its peak location $\beta_{\rm max}$
looks promising as a new criticality criterion.
Fig.\ \ref{Csusextra} illustrates its thermodynamic extrapolation:
we show a 3-parameter fit of the data points to the function $a/L^{c}+b$
(Fit 1); it leads $b=1.0959(62)$, with a good fitting quality of
$\chi^{2}/{\rm dof} = 1.36$,
which misses the consensus of the literature \cite{Has05},
$\beta_{\rm c} = 1.1199(1)$, by $2.14\, \%$, or $3.87 \sigma$.
We add another fit to the
same function, which includes $\beta_{\rm c}$ (Fit 2): it still has a
decent quality, with the ratio $\chi^{2}/{\rm dof} = 2.65$, which
means that it is still conceivable the $\beta_{\rm max}$ converges
to $\beta_{\rm c}$ in the large-$L$ limit. We are in the process
of testing this hypothesis with simulations on larger lattices.

\section{Final remarks}

We have reviewed some aspects of the BKT transition in the 2d XY
model, and we arrived at the recently suggested concept of semi-vortices.
We search for a new quantity which clearly detects the critical
temperature $T_{\rm c}$. The cluster vorticity susceptibility
$\chi_{\cal V}$ of eq.\ (\ref{clusus}) looks promising: it has a peak
close to $T_{\rm c}$, with a height that rises with the volume $L^{2}$.

So far it seems that a natural thermodynamic extrapolation of the peak
temperature --- based on data up to $L=264$ --- slightly misses $T_{\rm c}$.
However, we should keep in mind that finite-size effects tend to be 
very persistent in this model, in particular for the standard Hamiltonian
(\ref{Hami}). We saw this property in the case of the helicity modulus, and
the same holds for the magnetic susceptibility $\chi_{m}$:
at $T_{\rm c}$ it is predicted to scale as
$\chi_{m} \propto L^{7/4} (\ln L)^{1/8}$ \cite{Kos74}, but even size
$L=2048$ is not sufficient to numerically confirm these exponents
\cite{Has05}.

Hence, we hope for the large-$L$ peak location of $\chi_{\cal V}$ to
agree with $T_{\rm c}$, which would provide a new criterion for the
BKT transition.\\
  
\noindent
{\bf Acknowledgments:} This article is based on contributions by
BSGB and BDJH to the {\it XXXV Reuni\'{o}n Anual de la Divisi\'{o}n
de Part\'{\i}culas y Campos} and the {\it LXIV Congreso Nacional de
F\'{\i}sica}, respectively. Both were events of the {\it Sociedad Mexicana
  de F\'{\i}sica}; we thank the organizers. We are deeply indebted to
 Jo\~{a}o Pinto Barros for testing our results with an independent code.
 WB further thanks Stephan Caspar, Manes Hornung and Uwe-Jens Wiese for
 instructive discussions, and Raghav Jha for a helpful comment.
 The simulations were performed on the cluster of ICN-UNAM.
This work was supported by UNAM-DGAPA through PAPIIT project IG100219.

\end{multicols}
\medline
\begin{multicols}{2}

\end{multicols}
\end{document}